\pdfoutput=1 
\documentclass{aa}
\usepackage{graphicx}
\def\lunits{$\rm erg\,s^{-1}$~}
\def\funits{$\rm erg\,cm^{-2}\,s^{-1}$~}
\def\cunits{$\rm cm^{-2}~$}

\begin{document}
\title{X-ray selected Infrared Excess AGN in the Chandra Deep Fields: a moderate fraction 
 of Compton-thick sources.}


  \titlerunning{X-ray DOGs in the CDFs}
    \authorrunning{I. Georgantopoulos et al.}

   \author{I. Georgantopoulos\inst{1,2}
           E. Rovilos \inst{3}
           E. M. Xilouris \inst{2}
           A. Comastri\inst{1} 
           A. Akylas \inst{2}
           }

   \offprints{I. Georgantopoulos, \email{ioannis.georgantopoulos@oabo.inaf.it}}

   \institute{INAF-Osservatorio Astronomico di Bologna, Via Ranzani 1, 40127, Italy \\
              \and 
              Institute of Astronomy \& Astrophysics,
              National Observatory of Athens, 
 	      Palaia Penteli, 15236, Athens, Greece \\
              \and
         Max Planck Institut f\"{u}r Extraterestrische Physik,
 Garching bei M\"{u}nchen, Germany  \\
             }

   \date{Received ; accepted }

\abstract{We examine the properties of the X-ray detected, 
  Infrared Excess AGN   
 or Dust Obscured Galaxies (DOGs) in the Chandra Deep Fields (CDF). 
 We find 26  X-ray selected sources which obey the 24$\mu m$ to R-band 
 flux ratio criterion $f_{24}/f_R>1000$. These are at a median redshift of 2.3 while 
 their IR luminosities are above $10^{12}$ $\rm L_\odot$.
 Their X-ray luminosities are all above a few times $10^{42}$ \lunits 
 in the 2-10 keV band unambiguously arguing that these host AGN.  
  Nevertheless, their IR Spectral Energy Distributions are split 
 between AGN (Mrk231) and star-forming templates (Arp220).  
 Our primary goal is to examine their individual X-ray spectra 
 in order to assess whether this X-ray detected DOG 
 population contains heavily obscured or even 
 Compton-thick sources. 
  The X-ray spectroscopy reveals a mixed bag of objects.  
 We find that four out of the 
 12 sources with adequate photon statistics and hence 
 reliable X-ray spectra, show evidence for a hard 
 X-ray spectral index with $\Gamma\sim 1$ or harder,
 consistent with a Compton-thick spectrum. 
 In total 12 out of the 26 DOGs show evidence for flat spectral indices.
 However, owing to the limited photon statistics
  we cannot differentiate whether these are flat
 because they are reflection-dominated or because they show moderate 
 amounts of absorption. Seven DOGs  
  show relatively steep spectra $\Gamma > 1.4$
 indicative of small column densities.
 All the above suggest a fraction of Compton-thick sources that does not exceed  50\%.      
  The average X-ray spectrum of all 26 DOGs is hard ($\Gamma\sim  1.1 $) 
 or even harder ($\Gamma\sim0.6$) when we exclude the brightest sources. 
 These spectral indices are well in agreement with the stacked spectrum 
 of X-ray undetected sources ($\Gamma \approx 0.8$ in the CDFN).
 This could suggest (but not necessarily prove) that X-ray undetected DOGs, 
in a similar fashion to the X-ray detected ones presented here, 
  are hosting a moderate fraction  of Compton-thick sources.
     \keywords {X-rays: general; X-rays: diffuse emission;
X-rays: galaxies; Infrared: galaxies}}
   \maketitle
%

\section{Introduction} 

The hard X-rays (2-10 keV) present the advantage that they can penetrate large amounts of
  interstellar gas and thus can detect AGN which would be missed in optical wavelengths
 (Brandt \&  Hasinger 2005). 
 The Chandra mission has probed at unparallelled depth the X-ray Universe.
 The deep 2Ms observations in the Chandra Deep Field (CDF) North and South 
 (Alexander et al. 2003, Giacconi et al. 2002, 
 Luo et al. 2008) resolved 80-90\% of the extragalactic 
X-ray light, the X-ray background, in the 2-10 keV band revealing a sky density 
 of about 5000 
sources per square degree down to a flux of $1\times 10^{-16}$ \funits (Bauer et al. 2004). 
These deep CDF observations in combination with shallower 
 Chandra and XMM surveys (e.g. COSMOS, Hasinger et al. 2007;  AEGIS,
 Nandra et al. 2007) have provided the opportunity to study  
 with unprecedented accuracy the accretion history of 
 the Universe (Ueda et al. 2003, LaFranca et al. 2005, Barger et al. 2005, 
 Silverman et al. 2008, Brusa et al. 2009, Aird et al. 2010).    

However, even the extremely efficient X-ray surveys may be missing a fraction of 
heavily obscured sources. This is because at very high obscuring 
 column densities ($>10^{24}$ \cunits) 
around the nucleus even the hard X-rays (2-10 keV) are significantly suppressed. These are 
 the so-called Compton-thick AGN (see Comastri 2004 for a review) 
where the probability for Thomson scattering becomes significant.
  The X-ray background 
synthesis models (Comastri et al. 1995, Gilli et al. 2007) can explain the 
 peak of the X-ray 
 background between 20-30 keV, where most of its energy 
density lies, (eg Frontera et al. 2007, Churazov et al. 2007, Moretti et al. 2009) 
 only by invoking a numerous population of Compton-thick sources. 
 Nevertheless, the exact surface density of Compton-thick AGN required is still debatable 
 (see e.g. Sazonov et al. 2008, Treister et al. 2009). 
Additional evidence for the presence of an appreciable Compton-thick 
population comes from the directly measured space density of black holes in the 
 local Universe. It 
is found that this space density is a factor of 1.5-2  higher than that 
 predicted from the X-ray 
luminosity function (Marconi et al. 2004, Merloni \& Heinz 2008), although 
 the exact figure depends on the assumed efficiency in the conversion of 
  gravitational energy to radiation. This immediately suggests that 
 the X-ray luminosity function is missing a large number of AGN.    
According to the X-ray background models a small number of Compton-thick AGN should be
 lurking among the faint sources in the Chandra Deep Fields. 
As Compton-thick  sources have a quite distinctive X-ray spectrum in the 2-10 keV band, 
i.e. either a spectral turnover 
 in the transmission dominated case or a flat continuum in the reflection-dominated case, 
  X-ray spectroscopy provides  a reasonably robust way for identifying such 
heavily obscured sources. Tozzi et al. (2006) and  
 Georgantopoulos et al. (2007, 2009) have looked for these directly in the Chandra Deep Fields  
 using X-ray spectroscopy.   
 
  The advent of the IR {\it Spitzer} mission brought an additional perspective on the study   
  of Compton-thick sources.  
 This is because the absorbed optical and UV radiation heats the dust and is re-emitted 
  at IR wavelengths. This implies that Compton-thick sources should emit copious amounts 
 of mid-IR radiation.
 In particular, methods which combine the use of mid-IR and optical photometry 
 have been quite fruitful. 
  For example, the method proposed by Daddi et al. (2007) involves the 
selection of UV sources with mid-IR excess.
 Martinez-Sansigre et al. (2005) argue that a population of bright 24$\mu m$ AGN with no 
  3.6$\mu m$ detections is as numerous as unobscured QSOs at high redshift $z>2$.  
 Along these lines, Houck et al. (2005) have detected a population of 24$\mu m$ bright sources which are 
 very faint in the R-band  having $f_{24\mu m}/f_{R}>1000$.
 Dey et al. (2008) and Pope et al. (2008) argue that such sources are 
associated with Dust Obscured 
 Galaxies (DOG) at high redshift ($z\approx2$ with a small scatter $\sigma_z=0.5$). 
 At these distances the implied total IR luminosities are $\rm L_{IR} > 10^{12-14} L_{\odot}$
  comparable or in excess of low redshift Ultra-Luminous-Infrared galaxies, ULIRG
 (Mirabel \& Sanders 1996).  
  Theoretical simulations prove that DOGs  represent extreme gas-rich mergers 
 in massive halos with $\sim 10^{13} M_\odot$ (Narayanan et al. 2009).   
 Hereafter, we will adopt the notation 
 (DOGs) for mid-IR bright optically faint sources. 
  Fiore et al. (2008) argue that most DOGs may be associated 
     with heavily obscured sources
      {\it below the flux limit} of the 1Ms CDF-S survey. 
  The stacked X-ray 
       signal of the undetected DOGs in X-ray surveys, appears to be flat indicative 
        of absorbed sources, see    
         Fiore et al. (2008), Georgantopoulos et al. (2008), Fiore et al. (2009),
 Treister et al. (2009b),  Eckart  et al. (2010), Donley et al. (2010).
    In particular, Fiore et al. (2009) argue that the content of Compton-thick 
 sources among DOGs brighter than $550 mJy$ at 24$\mu m$ may be as high as 90\%.     
  Fiore et al. (2009) point out that at the bright mid-IR luminosities
    have been selected to ensure that any undetected X-ray source should 
     have a very low $L_X/L_{24 \mu m}$ ratio being either 
     a Compton-thick AGN or a galaxy.  
 Georgantopoulos et al. (2008) note that although it is possible that
 a significant number of Compton-thick sources lie among the X-ray undetected DOGs, 
  in the {\it Chandra} deep  fields 
it  is also likely that Low-Luminosity AGN with moderate absorption 
 may mimic a flat spectrum.   
         Finally, Pope et al. (2008) on the basis of {\it Spitzer} IRS spectroscopy, caution that the 
 normal galaxy (non-AGN) content of DOG samples may still be significant.  
            In any case, it is impossible to argue unambiguously that these sources 
             are Compton-thick given that only a stacked hardness ratio is available
              and not individual good quality X-ray spectra.             

 Here instead, we attempt to address this issue by focusing on the DOGs  
 which are   present among the  {\it detected}  sources in deep X-ray surveys.
 We examine their X-ray spectra as well as their mid-IR properties and we compare 
 with the properties of non X-ray detected DOGs. The primary aim is to examine 
 whether there is evidence for Compton-thick or at least heavily obscured sources among
 the X-ray DOGs.   
    We adopt $\rm H_o
=  75  \,  km  \,  s^{-1}  \,  Mpc^{-1}$,  $\rm \Omega_{M}  =  0.3$,\
$\Omega_\Lambda = 0.7$ throughout the paper.

\section{Data}
 
 \subsection{X-ray Data}

\subsubsection{CDF-N}
The CDF-N is centred at $\alpha = 12^h 36^m
49^s.4$,  $\delta =  +62^\circ 12^{\prime}  58^{\prime\prime}$ (J2000)
 The 2Ms  {\it Chandra} survey of  the CDF-N consists  of 20 individual
ACIS-I (Advanced CCD  Imaging Spectrometer) pointings observed between
1999 and 2002.  The combined observations cover a total area of $447.8
\,  \rm  arcmin^2$ and  provide  the  deepest  X-ray sample  currently
available together with the Chandra Deep Field South (Luo et al. 2008).   
  Here,   we   use    the   X-ray   source   catalogue   of
 (Alexander et al. 2003), which  consists of  503 sources detected  in at
least one of  the seven X-ray spectral bands  defined by these authors
in the  range $0.3-10$\,keV.  The flux  limit in the
$2-10$\,keV   band is $1.4\times
10^{-16}$\,\funits.  The Galactic column density towards
the CDF-N is $1.6\times 10^{20}$\,\cunits (Dickey \& Lockman 1990).

\subsubsection{CDF-S} 
The 2Ms CDF-S observations consist of 23 pointings. The first 1Ms
 (11 observations) was observed between 1999 and 2000. The analysis 
 of the 1Ms data is presented 
 in Giacconi et al. (2002) and Alexander et al. (2003).
 The analysis of all 23 observations is presented in 
 Luo et al. (2008).  
 The average aim point is $\alpha = 03^h 32^m
28^s.8$,  $\delta =  -27^\circ 48^{\prime}  23^{\prime\prime}$ (J2000).    
 The  23 observations cover an area of 435.6 arcmin$^2$ while  
 462 X-ray sources have been detected 
 by Luo et al. (2008).  
 The CDF-S survey reaches a sensitivity limit of $1.3\times 10^{-16}$ 
 \funits in the hard 
 2-8 keV band. 
 The Galactic column density towards
the CDF-S is $0.9\times 10^{20}$\,\cunits (Dickey \& Lockman 1990).

\subsection{Spitzer} 
The central regions of both the CDF-N and the CDF-S have been observed in
the mid-IR by the {\it Spitzer} mission (Werner 2000) as part of the Great
Observatory Origin Deep Survey (GOODS). These observations cover areas of
about $10 \times 16.5\rm \, arcmin^2$ in both fields using the IRAC (3.6, 4.5,
5.8 and  8.0$\, \rm \mu m$) and the  MIPS ($24\,  \rm  \mu m$) instruments onboard
Spitzer. Typical sensitivities for both surveys are $0.3\,\mu{\rm Jy}$ and
$80\,\mu{\rm Jy}$ for the $3.6\,\mu{\rm m}$ IRAC and $24\,\mu{\rm m}$ MIPS
respectively. The data products are available from the {\it Spitzer} data centre
({http://data.spitzer.caltech.edu/popular/goods/}).

\section{Optical Data}
Both GOODS-North and GOODS-South have been imaged with ground-based telescopes
as well as the {\it HST}. The {\it HST} observations are part of the
GOODS Legacy survey and are described in detail in Giavalisco et al. (2004). The
area imaged covers $2\times 180\,{\rm arcmin^2}$ with four ACS filters
(F435W, F606W, F775W, and F850LP). The typical sensitivity is 27.8\,mag(AB)
in the F850LP filter. 
 From the ground, an area larger than the GOODS-North has been imaged with the
KPNO-4\,m-MOSAIC and the Subaru-Suprime camera Capak et al. (2004) in the
U, B, V, R, I, and z' bands. The catalogue contains sources detected in the
R band with a limiting magnitude of 26.5(AB). The GOODS-South has been imaged
with the CTIO-4\,m-MOSAIC\,II camera as part of the MUSYC project
Gawiser et al. (2006). The survey is complete to a total magnitude of R=25(AB).

\section{Sample Definition}
We use the ``likelihood ratio'' method of Sutherland \& Saunders (1992) to
cross-correlate the positions of the sources in different catalogues. For the
northern field we follow the procedure described in detail in Rovilos et al.
 (2010). In short, we first find infrared (IRAC-$3.6\mu{\rm m}$)
counterparts to the X-ray sources and then we search the optical catalogue of
Capak et al. (2004) for optical identifications. {\it Spitzer}-MIPS counterparts
are again pinned to $3.6\,\mu{\rm m}$ positions. In order to select DOG
candidates we search for sources with $\log(f_{\rm 24\mu m}/f_R)>3$ defining
R=26.5(AB) as the lower limit for optical 'non-detections' and find 19 candidates.
We then inspect by eye the optical images of the 'non-detections' (15/19) to
check if there are any X-ray - IRAC sources with an obvious bright optical counterpart
which is not listed in the Capak et al. (2004) catalogue. We find 6 such cases.
We also remove source CDFN-109 from the list, because its MIPS counterpart is
blended with a nearby source (CDFN-107) and the $24\,\mu{\rm m}$ flux is not
reliable. The final northern sample of DOGs is listed in Table 1.

In CDF-S,  Grazian et al. (2006) provide the GOODS-MUSIC catalog. 
 This contains near-UV (U-band) magnitudes from the WFI at La Silla
 Arnouts et al. (2001) and VLT-VIMOS, optical magnitudes from the {\it HST}-ACS
 Giavalisco et al. (2004), near-infrared magnitudes ($J-H-K_S$) from VLT-ISAAC,
as well as mid-infrared magnitudes from {\it Spitzer}-IRAC. The multi-band
photometry has not been made by cross-correlating the various catalogues, but
with implementing the ``ConvPhot'' algorithm  de Santis et al. (2007) to perform
PSF-matching of the various images, and detecting the sources in the image with
the best quality, in this case the ACS-F850LP.

We have used the Grazian et al. (2006) catalogue and cross-correlated it with the
X-ray catalogue of  Luo et al. (2008, 2010) using the maximum likelihood method, and
also with the MUSYC catalogue to have a more complete coverage of the optical
bands. The MUSYC catalogue contains sources detected in the combined $B-V-R$
band and the $K_S$ band. We searched for
{\it Spitzer}-MIPS counterparts to the optical sources and have selected the
DOG candidates using the MUSYC R-band to compare with the $24\,\mu{\rm m}$ flux.
We found 22 candidate sources assuming a lower limit of 27.0\,mag(AB) for MUSYC
non-detections. We again optically inspected the MUSYC images of the
non-detections (11/21) for obvious bright optical counterparts not in the MUSYC
catalogue and found eight such  cases. The final southern sample of DOGs is listed in
Table \ref{sample}.

\begin{table*}
\caption{Xray, optical and infrared fluxes of the DOG sample} 
\label{sample}
\centering
\begin{tabular}{ccccccc}
\hline\hline
 ID      & RA (X-rays) & DEC (X-rays) & $f_{0.5-10\,{\rm keV}}$   & $R$ & $3.6\,\mu{\rm m}$ & $f_{24\,\mu{\rm m}}$ \\
  \hline
CDFN-92  & 189.0493 &  62.1707 &  2.50 &    25.967 &    17.834 &  160.6 \\
CDFN-129 & 189.0914 &  62.2677 &  1.22 &  $>$26.500 &    18.542 &  104.6 \\
CDFN-140 & 189.0986 &  62.1691 &  0.55 &  $>$26.500 &    18.820 &  148.2 \\
CDFN-167 & 189.1302 &  62.1659 &  0.21 &  $>$26.500 &    18.732 &  196.0 \\
CDFN-190 & 189.1483 &  62.2400 &  2.52 &     23.790 &    16.610 & 1426.0 \\
CDFN-220 & 189.1755 &  62.2254 &  0.28 &  $>$26.500 &    18.981 &  204.8 \\
CDFN-299 & 189.2413 &  62.3579 &  5.44 &  $>$26.500 &    18.077 &  199.1 \\
CDFN-302 & 189.2448 &  62.2498 &  0.12 &  $>$26.500 &    19.001 &  117.4 \\
CDFN-307 & 189.2472 &  62.3092 &  0.88 &  $>$26.500 &    18.815 &  271.0 \\
CDFN-317 & 189.2568 &  62.1962 &  0.09 &  $>$26.500 &    16.806 &  664.4 \\
CDFN-417 & 189.3563 &  62.2854 &  0.74 &  $>$26.500 &    18.380 &  172.6 \\
CDFN-423 & 189.3605 &  62.3408 &  1.40 &     25.159 &    17.394 &  909.5 \\
CDFS-95  &  53.0349 & -27.6796 &  4.29 &     25.586 & $<$25.986 &  241.0 \\
CDFS-117 &  53.0491 & -27.7745 &  2.47 &     26.366 &    22.047 &  110.0 \\
CDFS-170 &  53.0720 & -27.8189 &  0.13 &     26.704 &    21.692 &   83.4 \\
CDFS-197 &  53.0916 & -27.8532 &  0.59 &     26.116 &    22.547 &  238.0 \\
CDFS-199 &  53.0923 & -27.8032 &  0.52 &   $>$27.000 &    22.041 &  200.0 \\
CDFS-230 &  53.1052 & -27.8752 &  0.13 &     26.315 &    21.712 &  141.0 \\
CDFS-232 &  53.1070 & -27.7183 &  6.09 &     26.103 &    21.048 &  544.0 \\
CDFS-293 &  53.1394 & -27.8744 &  0.20 &  $>$27.000 &    22.959 &  351.0 \\
CDFS-307 &  53.1467 & -27.8883 &  2.51 &  $>$27.000 &    22.635 &   81.3 \\
CDFS-309 &  53.1488 & -27.8211 &  0.53 &     24.933 &    20.937 &  581.0 \\
CDFS-321 &  53.1573 & -27.8700 & 13.00 &     24.512 &    19.813 & 1080.0 \\
CDFS-326 &  53.1597 & -27.9313 &  0.81 &     26.759 &    21.353 &  177.0 \\
CDFS-346 &  53.1703 & -27.9297 &  7.26 &     24.582 &    21.528 &  717.0 \\
CDFS-397 &  53.2049 & -27.9180 &  2.11 &     27.009 &    21.155 &  163.0 \\
\hline\hline
\end{tabular}
\begin{list}{}{}
\item The columns  are: (1) Alexander ID number in the case of CDF-N or Luo ID 
 in the case of CDF-S (2) X-ray coordinates (J2000)  
 (3) total 0.5-10 keV flux in units of $10^{-15}$ \funits  
 (4) R-mag (AB) from Capak et al. (2004) (5) 3.6$\mu $ IRAC mag (6) 24$\mu m$ MIPS flux 
 in $\mu$Jy units  
\end{list}
\end{table*}

\section{X-ray Spectroscopy} 
The X-ray spectroscopy provides the most efficient tool for determining 
 whether an AGN is absorbed. In the case of moderate to heavy absorption 
  ($<10^{24}$ \cunits)  the energy of the absorption cut-off gives an accurate 
   measurement of the absorbing column density.
    On the other hand, the most heavily obscured sources,  
  the Compton-thick AGN, can be identified mainly through the
  presence of a very flat spectrum with a photon-index of $\Gamma\sim 1$.
   This flat spectrum is the signature of a reflected continuum from 
    obscuring material around the central source.    
 Thanks to the large exposure times, the 
  CDF observations yield a sufficient number of counts allowing for X-ray 
   spectroscopy to be performed in most cases.  

We use the {\sl SPECEXTRACT} script in the CIAO v4.2 software 
 package to extract spectra from the 20 individual CDF-N observations.
  The extraction radius varies between 2 and 4 arcsec with increasing off-axis angle. 
   At low off-axis angles ($<$4 arcmin)  this encircles 90\% of 
 the light at an energy of 1.5 keV.  
 The same script extracts response and auxiliary files. 
  The addition of the spectral, response and  auxiliary files has been 
   performed with the FTOOL  tasks {\sl MATHPHA}, {\sl ADDRMF} and {\sl ADDARF} respectively.  
    We use
the C-statistic technique (Cash 1979) specifically developed
to extract spectral information from data with low
signal-to-noise ratio. 
 We use the XSPEC v12.5 software
package for the spectral fits (Arnaud 1996).  

\subsection{Individual spectra}
We fit the data using a power-law
 absorbed by a cold absorber.  
 In the case where we have sufficient flux  
  ($f_x>1\times 10^{-15}$ \funits), we treat both the intrinsic column density 
and the photon index ($\Gamma_1$)  as free parameters. 
       In the rest of the cases where we have  limited photon statistics, we fix 
        the power-law photon index to $\Gamma_1=1.8$
 (Nandra \& Pounds 1994, Tozzi et al. 2006) 
         leaving only the column density as the free parameter. 
         The results are shown in Table 2.
 As it is customary in X-ray Astronomy, errors
 correspond to the 90\% confidence level. 
 
\begin{figure}
\rotatebox{270}{\includegraphics[width=7.0cm]{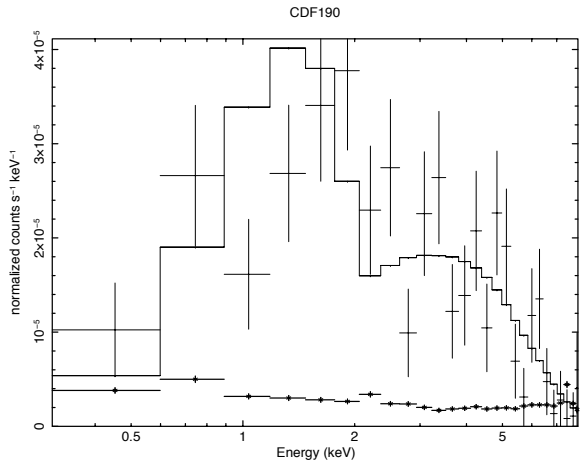}}
\caption{Power-law fit to the spectrum of the source CDFN-190. 
The background is also plotted as crosses}
\label{xspectra}
\end{figure}

\begin{figure}
   \begin{center}
\includegraphics[width=7.0cm]{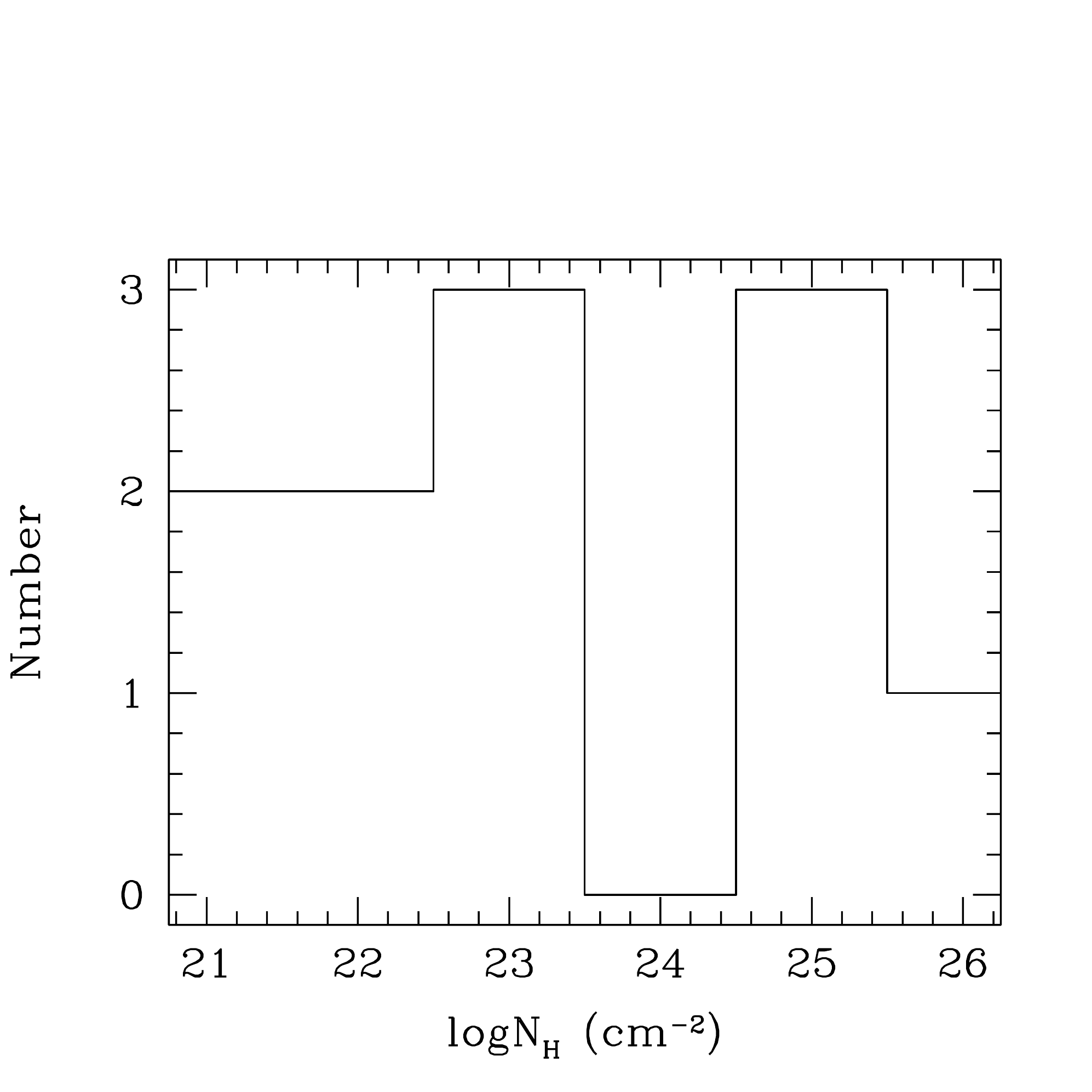} 
\includegraphics[width=7.0cm]{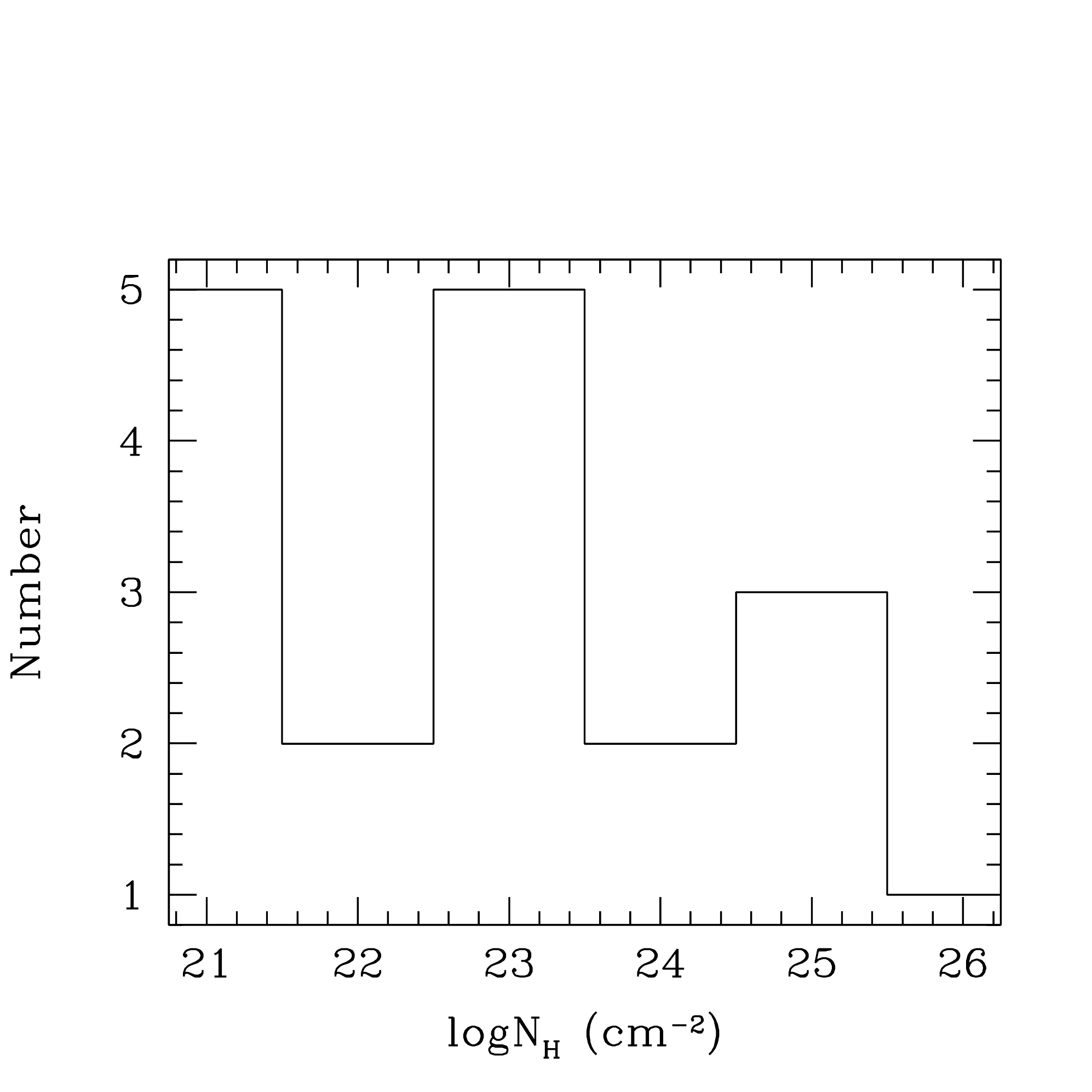}
\caption{Upper panel: rest-frame column density distribution of the brightest (11) sources
 with available redshift.
Lower panel: rest-frame column density for all (18) sources with available redshift.
Four sources with very low photon statistics have been also excluded.}
  \label{nh}
  \end{center}
\end{figure}

    First, we investigate the brightest sources i.e. those 12 with flux  
     $f_x  > 10^{-15}$ \funits.     
        There are four sources which  prefer a very flat power-law 
          index ($\Gamma_1 < 1 $ at the 90\% confidence level) 
 instead of intrinsic absorption: 92, 190, 423 in CDF-N,
  and 346 in CDF-S. 
  This flat spectrum is consistent with a reflection dominated X-ray emission  
 (e.g. Matt et al. 2004, Akylas \& Georgantopoulos 2009) and therefore these sources 
 can be considered as Compton-thick candidates.   
  We present an example X-ray spectrum (CDFN-190) in Fig. \ref{xspectra}.   
 We note that the properties of CDFN-92 and CDFN-190 have been 
             discussed in more detail in Georgantopoulos et al. (2009) 
            as candidate Compton-thick sources.
 We discuss their findings particularly in respect with the 
 presence of an FeK$_\alpha$ line.  A high equivalent width (EW) 
   FeK$_\alpha$ is often interpreted as the 'smoking gun' 
 giving away the presence of  a Compton-thick AGN. 
 CDFN-92 has an 90\% upper limit  of $\approx 0.7$ keV (rest-frame). 
  CDFN-190 which is associated with a sub-mm at a spectroscopic
 redshift of z=2.005 (Chapman et al. 2005), has a detected FeK$_\alpha$ line 
 with a rest-frame EW of $\sim 0.7$ keV.  
          CDFN-423 had not been included in the  
 sample of Georgantopoulos et al. (2009) as the source appeared 
 to be very hard but with a large error on the photon index. 
 The higher error had probably to do with the use of 
 different fitting statistics ($\chi^2$) in that paper
 in combination with the limited photon statistics
 ($f_{2-10}\approx 10^{-15}$ \funits). No FeK$_\alpha$ line is detected in the spectrum         
  of CDFN-423 with the 90\% upper limit on the rest-frame EW being 0.65 keV. 
 Finally,  no FeK$_\alpha$ line is detected in the spectrum of CDFS-346 at rest-frame energy 
 of 6.4 keV (360 eV upper limit). 
      The four flat sources constitute an appreciable fraction (30\%) 
       of the sources with good quality X-ray spectra.
        We show the rest-frame column density distribution of these  bright sources in Fig. \ref{nh}
         (excluding CDFS-95 for which there is no redshift).
         We count all the four flat spectrum sources as Compton-thick sources 
          and  therefore we assign a value of $\rm N_H=10^{24}$ \cunits for these at the 
           observer's frame.
            For simplicity, we consider all the column density upper limits as unabsorbed sources,
             assigning $\rm N_H<10^{22}$ \cunits.
  Moreover, we show the column density distribution for our full sample 
in  Fig. \ref{nh} (lower-panel): 18 spectra have been considered excluding the 
 four faintest sources where a spectral fit is not feasible as well as the four sources with no 
  redshift available. 
  
\begin{figure}
   \begin{center}
\includegraphics[width=8.5cm]{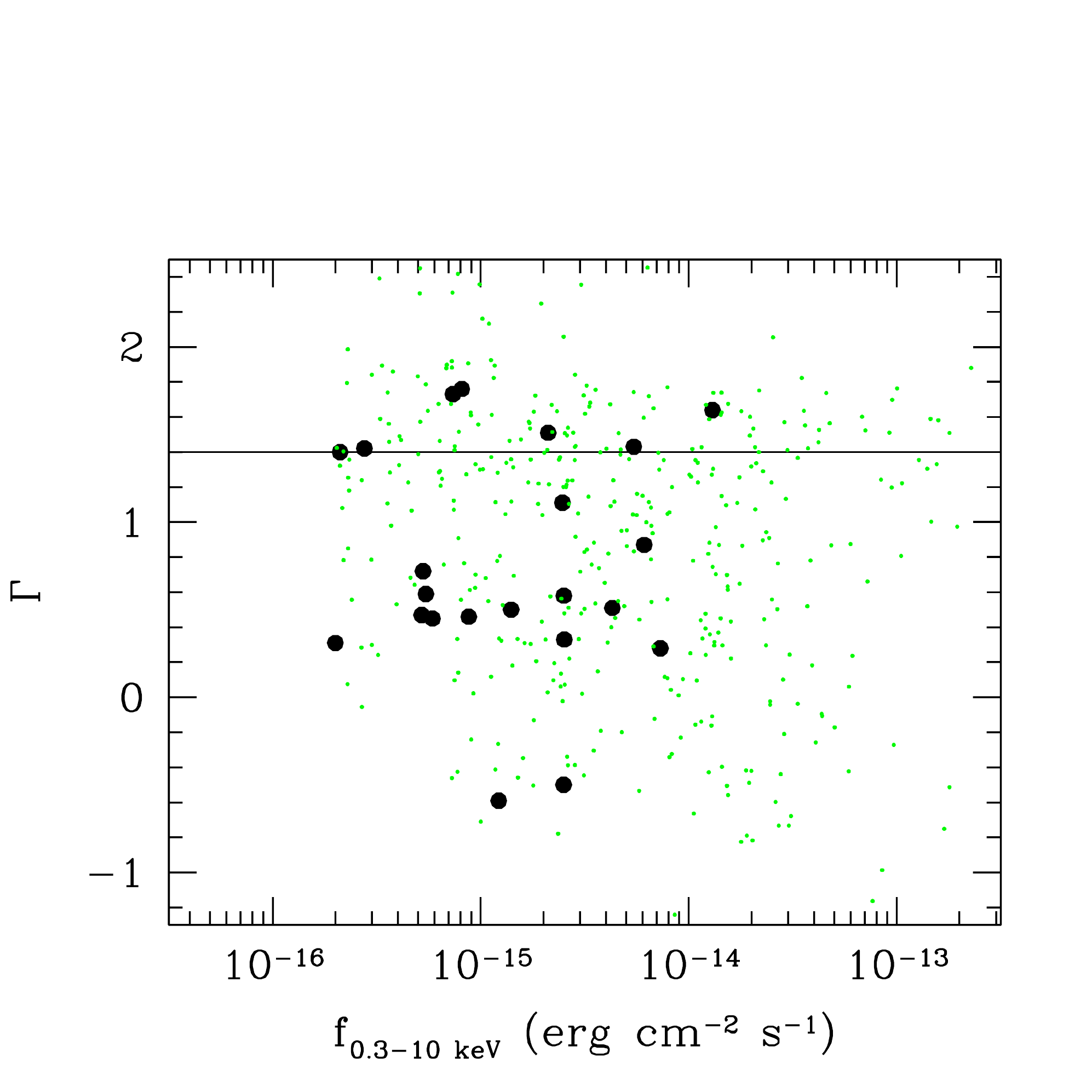} 
\caption{The Photon index $\Gamma_2$ of X-ray detected DOGs as a 
function of the total band flux
 (filled circles) for column density  fixed to $N_H=0$. For comparison 
 we plot the photon indices of all sources in the CDF-N (small dots).}
  \label{fxgamma}
  \end{center}
\end{figure}

          In addition, we fit all the spectra with the simplest
 possible model i.e. a  
           power-law model without absorption.
            The photon index  $\Gamma_2$ results are given again in Table 2.  
 This exercise is quite useful in demonstrating the effects of spectral 
 degeneracy. For example,  CDFS-117  with a steep spectral 
 index of $\Gamma_1=1.81$  and an observer's-frame column density of only 
 0.43$\times10^{22}$ $\rm cm^{-2}$ in the absorbed power-law model, 
 has a flat equivalent photon-index of $\Gamma_2\sim 1.1$, (in the power-law model), 
   which could have been interpreted as a 
 characteristic of a reflection dominated spectrum.     
       In Fig. \ref{fxgamma} we plot the photon index $\Gamma_2$ 
       as a function of the total X-ray flux.
             We see that  the DOGs span a large range 
              in both flux and photon index. 
   Seven sources are relatively unabsorbed having $\Gamma_2 >1.4$.  
  A photon index of $\Gamma =1.4$ corresponds to  the stacked (co-added) spectrum of all 
 sources in the CDF-S (Tozzi et al. 2001). The above
 figure has to be interpreted with caution because 
  of the effects of spectral degeneracy mentioned above.  
 From this plot, we can securely tell which sources are relatively steep 
 and thus unabsorbed i.e. those with $\Gamma_2 > 1.4$.
 Thus we are confident that 7 out of 22 sources present little absorption.  
 However, we cannot use this plot to 
  find Compton-thick sources.
 This is because the flat $\Gamma_2 $ sources could be
  hard because of an intrinsically flat spectral index 
 suggestive of a reflection spectrum
 but alternatively because of just moderate absorption.  
 Only by examining sources which have 
 adequate photon statistics and by 
 leaving both the photon index and the 
 column density as  free parameters 
 we can have some indication on 
 whether a source is truly intrinsically flat. 
 Inspection of the photon index $\Gamma_2$ column 
 in Table 2 shows that there are 12 sources 
 which appear flatter than $\Gamma<1$. In this calculation 
 we excluded sources CDFS-232 and CDFS-307 in which 
 a fit with a moderate absorbing column is preferable.

\begin{table*}
\begin{center}
\caption{X-ray spectral fits}
\label{xtable}
\begin{tabular}{lccccc}
\hline 
ID  & z &  $N_H$      & $\Gamma_1$       & $\Gamma_2$  & $L_x$   \\ 
(1)   & (2)&    (3)        &  (4)          &  (5)         &   (6)       \\
\hline
  CDFN-92$^\dagger$ & 0.95 & $<$1  & $-0.45^{+0.35}_{-0.40}$   & $-0.50^{+0.34}_{-0.37}$ &  0.28  \\
 CDFN-129 & 2.33 & $ <3.5$ & $0.29^{+1.27}_{-1.0}$  & $-0.59^{+0.46}_{-0.51}$ & 4.8  \\
 CDFN-140 &  -    & $0.8^{+0.5}_{-0.34}$          &    1.8          & $0.59^{+0.37}_{-0.37}$  & 0.41  \\
 CDFN-167 & -    & $0.13^{+0.27}_{-0.13}$ &   1.8    &  $1.40^{+0.55}_{-0.52}$    & 0.46 \\
 CDFN-190$^\dagger$  & 2.005 & $<0.04$ & $0.33^{+0.17}_{-0.21}$ &  $0.33^{+0.17}_{-0.21}$   & 1.28  \\
 CDFN-220 & 2.34 & $<0.4$  & 1.8                 &  $1.42^{+0.78}_{-0.65}$   & 0.3  \\
 CDFN-299 & 2.36 & $0.12^{+0.04}_{-0.03}$ & $1.74^{+0.20}_{-0.20}$         &  $1.43^{+0.11}_{-0.10}$ & 104.  \\ 
 CDFN-302 & 1.59 &  -         &  1.8        &     1.8 &  0.37 \\  
 CDFN-307 & -    & $1.2^{+0.4}_{-0.4}$   & 1.8                &   $0.46^{+0.27}_{-0.44}$ & 1.39  \\ 
 CDFN-317 & 2.28  & - & 1.8          & 1.8  & 0.27 \\ 
 CDFN-417 & 1.90 & $<0.13$ &    1.8           & $1.73^{+0.25}_{-0.25}$ & 3.0 \\       
 CDFN-423$^\dagger$  & 2.07 & $<0.16$  & $0.48^{+0.4}_{-0.4}$   & $0.50^{+0.45}_{-0.40}$ & 12.0  \\
 CDFS-95    & -    & $<0.80$            & $1.00^{+0.72}_{-0.74}$    & $0.51^{+0.36}_{-0.39}$ & 1.5  \\
 CDFS-117 & 2.24 & $0.43^{+0.17}_{-0.20}$ & $1.81^{+0.23}_{-0.34}$   & $1.11^{+0.18}_{-0.16}$ & 2.56 \\
 CDFS-170 & 2.25 & -                  & -                      & 1.8                    & 0.3    \\
 CDFS-197 & 2.57 & $1.3^{+1.0}_{-0.60}$ & $1.8$                   & $0.45^{+0.56}_{-0.61}$ & 0.4 \\
 CDFS-199 & 2.50 &  $0.77^{+0.70}_{-0.45}$                 &  1.8                      & $0.47^{+0.51}_{-0.58}$ & 0.6 \\ 
 CDFS-230  & 2.73 & -                 &  -                      &  1.8                 & 0.7    \\
 CDFS-232 & 2.291 & $0.62^{+0.16}_{-0.16}$ & $1.77^{+0.13}_{-0.24}$ &  $0.87^{+0.12}_{-0.10}$ & 5.6  \\
 CDFS-293 & 3.62 &  $0.8^{+0.8}_{-0.4}$                  &   1.8                 &  $0.31^{+0.53}_{-0.43}$                  & 1.4 \\
 CDFS-307 & 1.83 & $1.0^{+0.5}_{-0.4}$  &  $1.67^{+0.26}_{-0.38}$ & $0.58^{+0.17}_{-0.20}$  & 1.6  \\
 CDFS-309 & 2.57 & $0.8^{+0.8}_{-0.6}$ & $1.8$    & $0.72^{+0.78}_{-0.72}$ & 0.4 \\
 CDFS-321 & 1.603 & $0.2^{+0.04}_{-0.04}$ & $2.2^{+0.12}_{-0.11}$ & $1.64^{+0.06}_{-0.06}$ & 9.0 \\
 CDFS-326 & 2.50  &  $<0.1$                 &  1.8                  & $1.76^{+0.46}_{-0.36}$ & 1.8 \\
 CDFS-346$^\dagger$  & 3.06  &  $0.47^{+0.30}_{-0.30}$ & $0.75^{+0.26}_{-0.34}$ & $0.28^{+0.14}_{-0.17}$ & 4.8  \\ 
 CDFS-397 &  1.56 & $<0.07$             & $1.47^{+0.32}_{-0.19}$ & $1.51^{+0.26}_{-0.22}$ & 1.2 \\
\hline
\hline 
\end{tabular}
\begin{list}{}{}
\item The columns  are: (1) X-ray ID as in Table 1 (2)  redshift;  three and two decimal numbers 
 refer to spectroscopic and photometric redshift respectively; (3) Intrinsic {\it Observer-frame}
 column density in  units 
 of $10^{22}$ $\rm cm^{-2}$ for the power-law+absorption model;  
 (4) Photon index in the case of the 
 power-law+absorption model; 
 (5) Photon index in the case of  the power-law only model;
 Note that values with no error bars imply that the parameters have been fixed to this value.
 In four cases (CDFN-302, CDFN-317, CDFS-170 and CDFS-230), 
 the very limited photon statistics did not allow even a 
 single free parameter fit and therefore $\Gamma_2$ has been fixed to 1.8. 
 (6) observed (uncorrected for absorption) luminosity in the 2-10 keV band in units of $10^{43}$ \lunits.
 In the case where there is no redshift available we assign the median redshift z=2.3.
  $^\dagger$  denotes candidate Compton-thick sources.      
\end{list}
\end{center}
\end{table*}

\subsection{Average X-ray spectrum}
We derive the average X-ray spectrum of our sources separately for the 
 CDF-N and CDF-S. 
  The spectra are fit jointly with the photon index tied to a common  value. 
 We use the C-statistic in the spectral fits.  
  The average spectra are fit   
   in the 0.3-4 keV band. This band has been selected by 
    Fiore et al. (2008) and Georgantopoulos et al. (2008) 
     for the derivation of the stacked  signal of non-X-ray  
      detected DOGs in the CDF-S and CDF-N respectively.  
   Their choice was driven mainly by the fact that at high energies  
   the effective area rapidly decreases while the particle background 
 becomes high.
  The best-fit results are shown in Table 3.
 The total number of counts in the spectra is given as well 
(source + background). As the total spectrum may be dominated by 
 the most bright source, we choose to present as well the best-fit 
 results  excluding the source which contains the most photon counts.
 In the CDF-N the co-added photon index is very hard     
 ($\Gamma \approx 1.07 $) while when we exclude the brightest source
  we obtain an even  harder spectrum with $\Gamma \sim 0.6$.
 In the CDF-S the spectrum is comparable, having $\Gamma \approx 1.13$
 and $\Gamma \sim 0.6$ when the brightest source is excluded. 
 All photon indices above are significantly harder than  
 $\Gamma \approx 1.4$ which corresponds to the co-added spectrum of all 
 sources in the CDF-S (Tozzi et al. 2001) as well as to the spectrum 
 of the X-ray background in the 2-10 keV band (De Luca \& Molendi 2004). 
 
 \begin{table*}
 \begin{center}
 \caption{Average X-ray spectrum}
\label{average} 
\begin{tabular}{llll}
 \hline 
  Sample           &No        &  Counts     & $\Gamma$ (0.3-4 keV) \\
  \hline
  CDF-N           & 12       &   1831            & $ 1.07\pm0.12$       \\
  CDF-N (excl. \#299)  & 11   &   1048             & $0.63\pm 0.17$      \\
  CDF-S                & 14     &  4105           & $1.13\pm 0.06$       \\
  CDF-S (excl. \#321)    &  13 &   2121             & $0.65\pm 0.09$   \\
  \hline
 \end{tabular}
 \end{center}
 \end{table*} 

We also derive the average spectrum separately for sources with power-law 
 (AGN) SEDs and star-forming type SEDs (see next section for details). 
The results are given in Table 4. 
 We see that there is no significant difference in the average X-ray spectrum of 
 the two types of populations, at least when the brightest source is excluded.  

\begin{table*}
 \begin{center}
 \caption{Average X-ray spectrum according to IR SED}
\label{average_sed} 
\begin{tabular}{llll}
 \hline 
  Sample           &No        &  Counts     & $\Gamma$ (0.3-4 keV) \\
  \hline
  Star-forming SED           &  19      &   2672            & $ 0.94\pm0.08$       \\
  Star-forming SED (excl. \#299)  & 18   &   1889             & $0.58\pm 0.10$      \\
  AGN SED                &   7   &   3264           & $1.22\pm 0.07$       \\
  AGN SED (excl. \#321)    & 6  &    1280            & $0.66\pm 0.11$   \\
  \hline
 \end{tabular}
 \end{center}
 \end{table*}

\begin{table*}
 \begin{center}
 \label{lir}
 \caption{Infrared Luminosities}
 \begin{tabular}{lllll}
 \hline 
  ID$^1$      & $ L_{IR}$$^2$            &  $\nu L_\nu(6\mu m)^3$  & $\log (L_{2-10}/L_{6\mu m})$$^4$  & Template$^5$ \\
 \hline
 CDFN-92$^\dagger$ & 2.2  & 0.11 & -1.19 & 1   \\
CDFN-129 &  6.0 & 0.29  & -0.69 & 1  \\
CDFN-140 & 3.0 & 0.15 &   -1.26   & 1 \\
CDFN-167 & 4.2 & 0.2 &  -1.2   & 1  \\
CDFN-190$^\dagger$ & 8.4  & 16.   &  -2.69 & 2 \\
CDFN-220 & 6.0 & 0.29    & -1.58  & 1 \\
CDFN-299 & 12. & 0.6 & 0.63 & 1 \\
CDFN-302 & 2.0 &0.1  & -1.02 & 1 \\
CDFN-307 & 1.1 & 2.0 &   -1.75   &  2 \\
CDFN-317 & 23. & 1.1  & -2.21 & 1 \\
CDFN-417 & 5.6 & 0.27 & -0.55 &  1 \\
CDFN-423$^\dagger$ & 4.9 & 9.2  & -1.49 & 2 \\
CDFS-95  & 11.7 & 0.57 &   -1.18  &  1 \\
 CDFS-117 & 0.8 & 1.51 & -1.36 & 2 \\
 CDFS-170  & 3.8 & 0.18 & -1.33 & 1 \\
 CDFS-197 & 4.75  & 0.23 & -1.36 & 1 \\
 CDFS-199 &   4.53  &  0.22   &   -1.16   &  1\\
 CDFS-230 & 7.8 & 0.38  &     -1.33             & 1 \\
 CDFS-232 & 3.9 & 0.73  & -1.71 &  2 \\     
 CDFS-293 &  4.2    & 0.20    & -0.96     & 1 \\
 CDFS-307 &  1.29    &  0.06   &   -0.20   & 1 \\
 CDFS-309 & 3.6  & 6.77 & -2.82 & 2  \\  
 CDFS-321 & 3.1  & 5.7 & -1.41 &  2 \\
 CDFS-326 &  3.26    &   0.16  & -0.55     &  1\\
 CDFS-346$^\dagger$ &  0.64    & 12.    &   -2.0   & 1 \\
 CDFS-397 & 5.6  & 0.28 & -0.97 & 1 \\ 
  \hline
 \end{tabular}
\begin{list}{}{}
\item The columns  are: (1) X-ray ID as in table 1
  (2) Total (8-1000 $\mu m $) IR luminosity in units of $10^{12}$ $L_\odot$ 
 (3) 6$\mu m$ $\nu L_\nu$ monochromatic luminosity in units of $10^{11}$ $L_\odot$ 
 (4) ratio of 2-10 keV X-ray  to $6\mu m$  monochromatic luminosity (where both luminosities in units \lunits)
  (5) Template used from Polletta et al. (2007); 1 corresponds to Arp220 while 2 to Mrk231. 
\end{list}
 \end{center}
 \end{table*} 

\begin{figure*}
   \begin{center}
\includegraphics[width=12.cm]{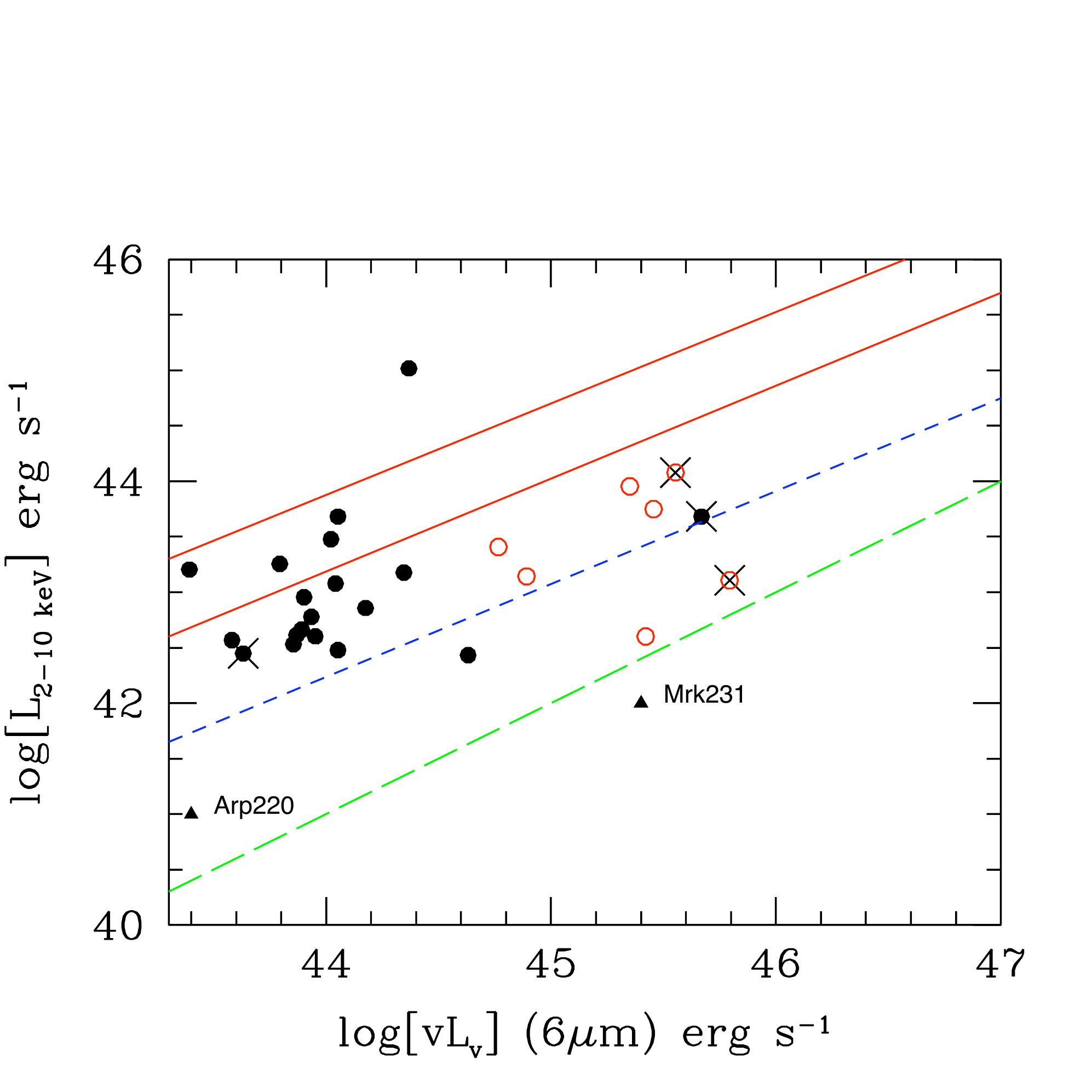} 
\caption{The 2-10 keV X-ray luminosity as a function of the 6$\mu m$ 
 luminosity. The black (filled) and red (open) circles denote 
 star-forming and AGN SEDs respectively. 
 The solid lines denote the region occupied by unabsorbed 
 AGN while the region beyond the short dash line is populated by Compton-thick 
 AGN in the local Universe (adapted from Bauer et al. 2010).
 The long dash line denotes the $\rm L_x/L_{6\mu m}\approx 10^{-3}$ luminosity ratio of the undetected
 DOGs in the CDF-N (from Georgakakis et al. 2010).  
 The crosses denote the flat-spectrum sources (see text).
 Finally for comparison we plot the positions of Mrk231 and Arp220
 on the diagram.}
  \label{lxl6}
  \end{center}
\end{figure*}

\begin{figure*}
   \begin{center}
\includegraphics[width=10.cm]{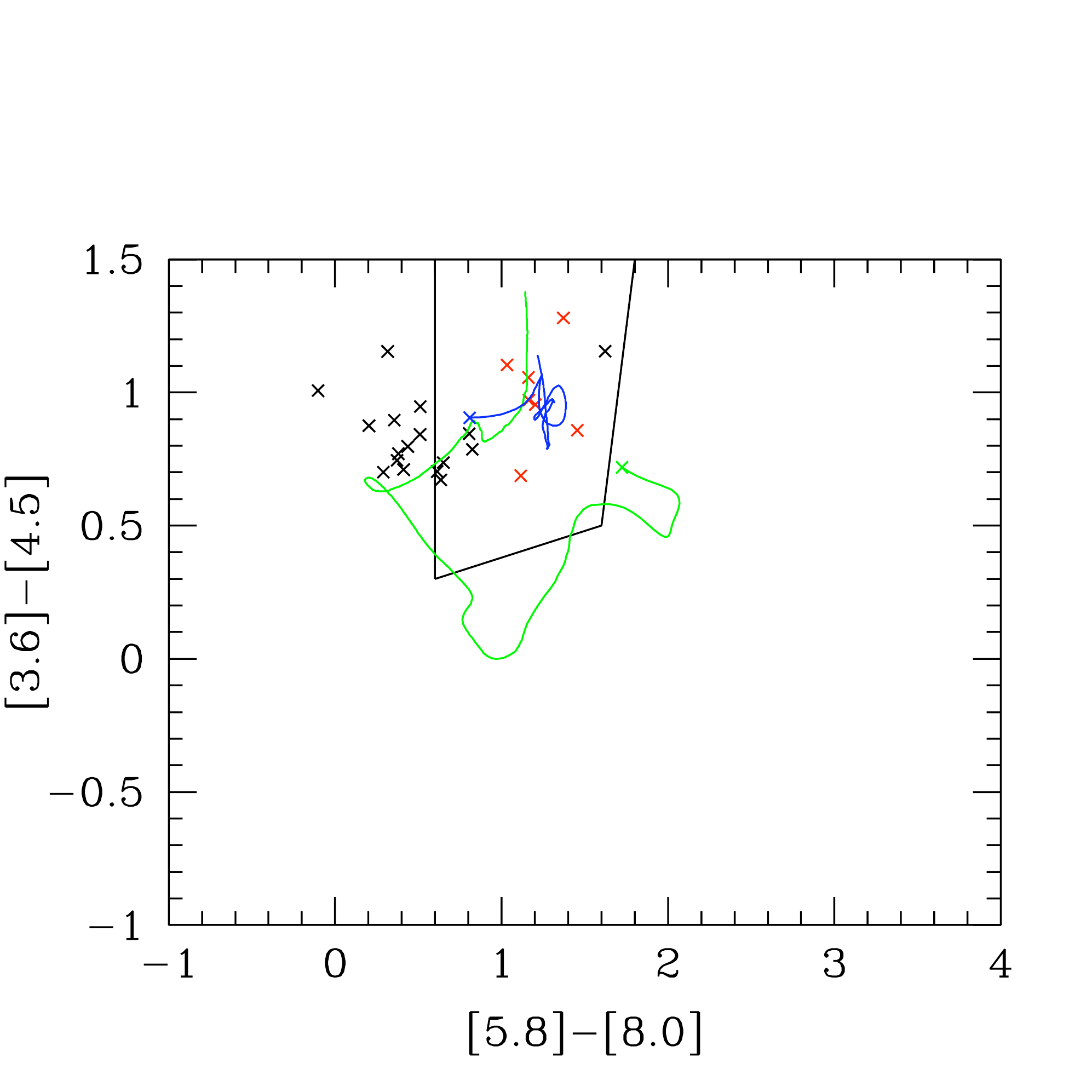} 
\caption{Mid-IR colour-colour plot. Black and red crosses 
 correspond to start-forming and AGN SEDs respectively. 
 The wedge defines the AGN region following 
the selection criteria of Stern et al. (2005). The colour tracks define two 
different template SEDs as function of redshift up to z = 3. 
The blue line represents the Mrk231  (ULIRG, Compton-thick) template,  while the 
 green line represents Arp220 (ULIRG, star-forming) SED. 
  For each template the position 
of the redshift z = 0 is marked with a large cross. The templates are 
 taken from Polletta et al. (2007).     
}
  \label{stern}
  \end{center}
\end{figure*}

\begin{figure}
   \begin{center}
\includegraphics[width=8.5cm]{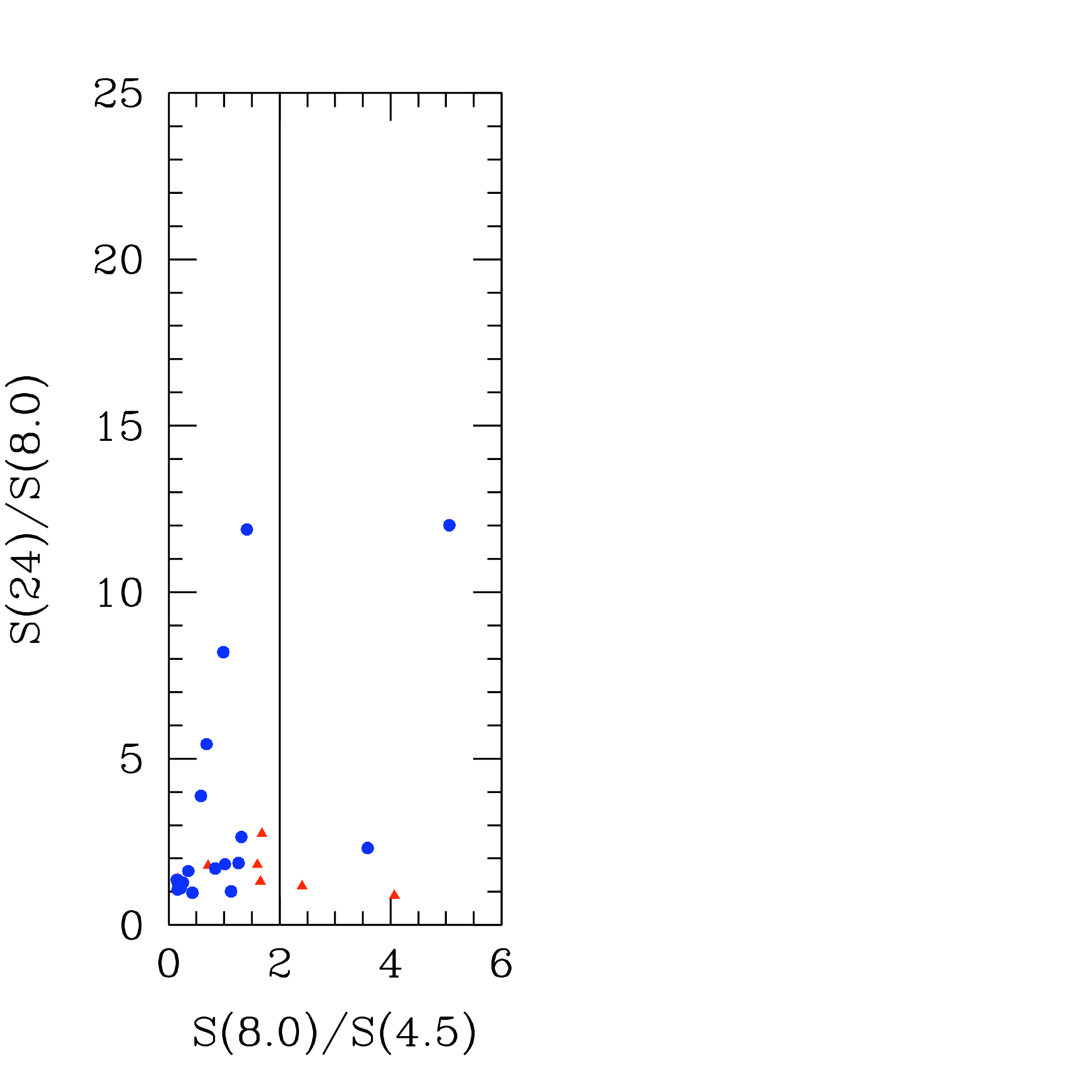} 
\caption{The Spitzer colour-colour diagram proposed to separate 
star-forming from AGN dominated DOGs. Circles (blue) and triangles (red) 
 represent the DOGs in our sample having star-forming and AGN SEDs respectively. The solid line denotes the 
 dividing line between AGN and star-forming galaxies at  
$S_{8.0}/S_{4.5}=2$}
  \label{pope}
  \end{center}
\end{figure}

\section{Mid-IR properties}

\subsection{Mid-IR Spectral Energy Distributions}
\begin{figure*}
   \begin{center}
\includegraphics[width=16.0cm]{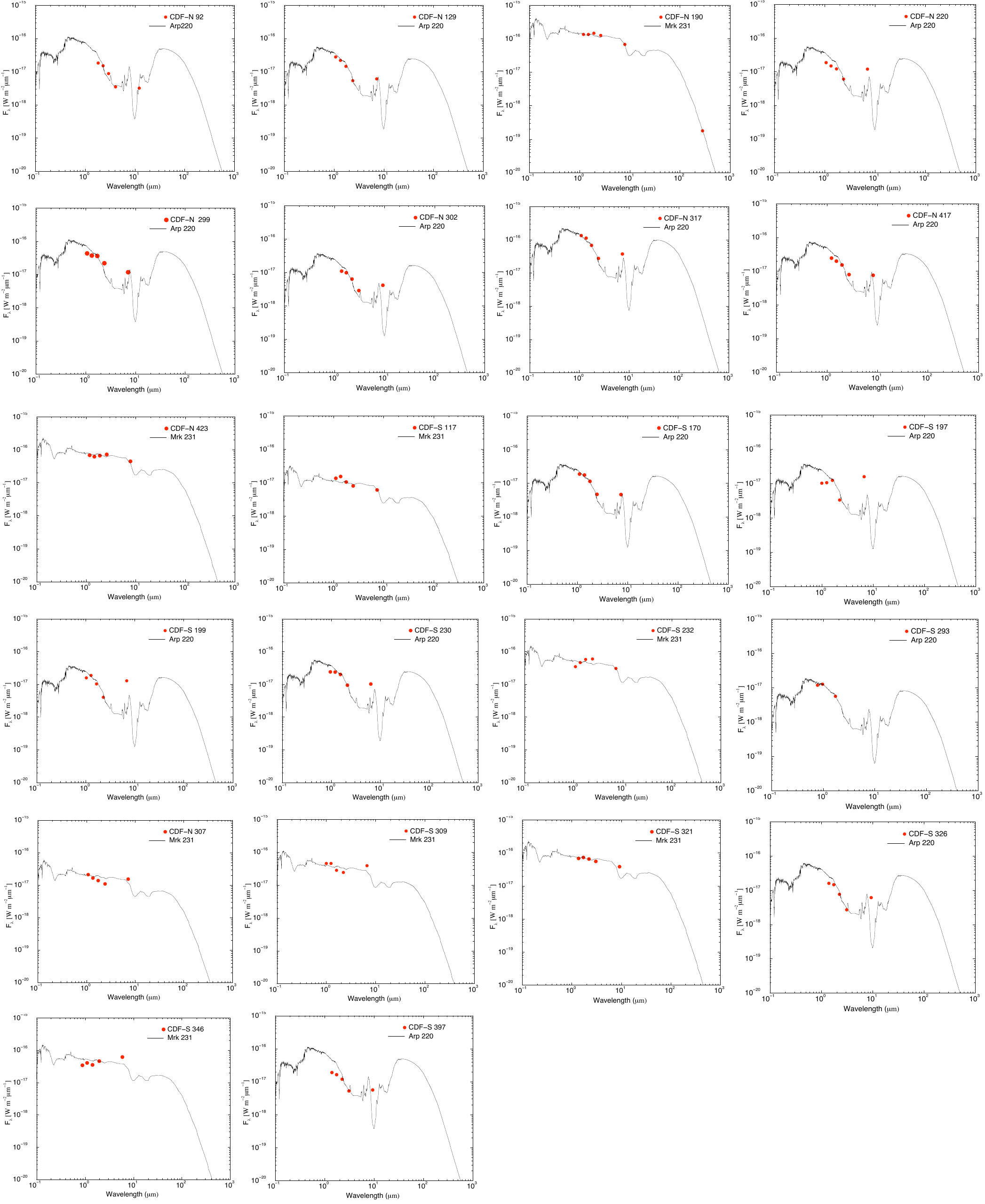} 
\caption{The IR Spectral Energy Distributions}
  \label{SED}
  \end{center}
\end{figure*}

 We construct the Spectral energy distribution (SED) 
 in order to estimate the IR luminosities
 and also to get some idea on the dominant 
 powering mechanism (AGN or star-formation) in the 
 mid-IR part of the spectrum.   
 We present the IRAC and MIPS $24\mu m$ SED 
 of the sources with available redshift 
 in Fig. \ref{SED}. For one source (CDFN-190) there are available 
 sub-mm data (Chapman et al. 2005).  
 We also overplot the best matching 
 SED template from Polletta et al. (2007).
  Polletta et al. (2006, 2007) 
 have made extensive studies of the SEDs of X-ray selected AGN.  
 Polletta et al. (2007) provide an inventory of template SEDs for various 
 types of AGN and star-forming  galaxies.    
 In many cases the template of Mrk231 provides a satisfactory 
 representation of the data suggesting an AGN underlying continuum. 
 Mrk231 is the nearest Broad-Absorption-Line QSO (Braito et al. 2004) and an 
 Ultra-Luminous-Infrared galaxy 
 (ULIRG). The SED of the remaining sources is well represented by the Arp220
 template. 
 These are the sources with a distinct dip in their mid-IR spectra at a
  rest-frame wavelength below 10$\mu m$ 
 Arp220 (e.g. Iwasawa et al. 2005) is a ULIRG 
  whose far-IR SED is dominated by a very strong star-forming component. 
 The IR luminosities are presented in table 5.         
 We emphasize that regardless of the mid-IR classification, 
 most of our 26 DOGs can be classified almost certainly as AGN 
 on the basis of their high X-ray luminosity and $L_X/L_{6\mu m}$ diagram
  (see below). 
 Their 2-10 keV X-ray luminosity (see Table 2)
 exceeds the $10^{42}$ \lunits limit which is considered 
 to be the dividing line  between star-forming 
 galaxies and AGN (e.g. Zezas, Georgantopoulos \& Ward 1998,  
 Ranalli, Comastri \& Setti 2003). 
 Therefore the term 'star-forming' mid-IR SED should not 
 be misleading.  

\subsection{X-ray to $6\mu m$ mid-IR luminosity ratio} 
  One of the most reliable proxies of the intrinsic power of an 
 AGN is the mid-IR luminosity (e.g. Lutz et al. 2004). 
 The mid-IR emission is believed to provide a good measurement of the 
 AGN luminosity  even in the most obscured cases, 
  since it is more extended than the X-ray absorbing material 
  and thus represents an isotropic AGN property.
  Consequently we use the observed X-ray to mid-IR luminosity to provide further 
   clues on whether our sources are obscured, expecting that the ratio of the X-ray 
    to the mid-IR luminosity should be suppressed in the most obscured sources 
     (eg Alexander et al. 2005, Alexander et al. 2008).
 Here, we use  the monochromatic $6\mu m$ mid-IR luminosity instead of the full IR
 luminosity. This is because the former is more representative 
 of the hot dust emission ($>300 K$) and thus provides a 
 better diagnostic of the AGN power.         
 The $6\mu m$ luminosity is derived from our SED fitting.   
  
 We present the  monochromatic 6$\mu m$ IR  luminosity
   against the 2-10 keV absorbed luminosity in Fig. \ref{lxl6}.
 The area between the solid lines denotes the region of 
  of the X-ray to 6$\mu m$ plane where local AGN reside (Lutz et al. 2004).
  The area below the dashed line corresponds to low X-ray 
 luminosity sources i.e. Compton-thick sources (or alternatively 
 normal galaxies).      
  The crosses denote the sources with flat spectral index $\Gamma$.
  We see a relatively good correspondence between 
 the flat-$\Gamma$ sources and the low $L_x/L_{6 \mu m}$ sources.
   One of the lowest $L_x/L_{6 \mu m}$ sources  is CDFN-190, a sub-mm galaxy 
 with a spectroscopic redshift of z=2.005 (Chapman et al. 2005).
  Other low $L_x/L_{6\mu m}$ sources include
  CDFN-317 and CDFS-309. 
 The first is among the two faintest 
 sources in the CDF-N ($f_x\sim 1 \times 10^{-16}$ \funits) where 
 the very poor photon statistics did not even allow a 
 single-parameter spectral fit ($L_x\approx 4\times 10^{42}$ \lunits). 
  The photon statistics of CDFS-309 allow a single parameter fit only.
 Its photon index is flat with $\Gamma_2=0.72^{+0.78}_{-0.72}$.
 Alternatively, assuming $\Gamma =1.8$ its {\it rest-frame}  column density
 would be $4\times 10^{23}$ $\rm cm^{-2}$.      

\subsection{IRAC Colour-Colour diagram} 
 
It is interesting to examine the position of our sources 
 on the IRAC colour-colour diagram. 
Mid-IR colour selection has been suggested as 
 a powerful tool for  selecting  AGN
 (see e.g Martinez-Sansigre et al. 2005).   
  The principle behind this approach is that luminous AGN 
have power-law spectral energy distribution in the mid- 
IR, while galaxies have characteristic black-body spectra that 
peak at about 1.6 $\mu m$. The mid-IR colours of AGN are therefore 
redder than those of galaxies defining a characteristic wedge. 
Different combinations of mid-IR colours have been proposed to 
select AGN (e.g. Lacy et al. 2004, Stern et al. 2005).
The Stern et al. (2005) criteria are likely to suffer less contamination from 
normal galaxies, while at the same time remain relatively sensitive to  
obscured AGN (but see Barmby et al. 2006 and Cardamone et al. 2008). 
   This method is graphically shown in Fig. \ref{stern}. 
 We see that all power-law SED fall within the wedge. However, there is also 
 an appreciable number of DOGs which fall leftwards of the wedge
 close to the region of the parameter space occupied by Arp220 at 
  high redshifts.  
 The same trend has been observed for the X-ray undetected DOGs in 
 the CDF-N (Georgantopoulos et al. 2008).

Pope et al. (2008) present IRS spectroscopy for  sample 
of 12 DOGs in the CDF-N. The spectroscopy shows that 
their sample is divided in half between 
AGN and star-forming galaxies. On the basis of their 
spectroscopic results these authors propose a colour diagram 
 for the selection of AGN.    They find that AGN-dominated 
  sources separate very nicely at $S_{8.0\mu m}/S{4.5 \mu m}=2 $.
   We show this diagnostic diagram for our sources in Fig. \ref{pope}. 
    Interestingly most of our sources fall in the star-forming 
     galaxy regime  although the vast majority are bona-fide AGN 
      on the basis of their high X-ray luminosity and high $L_X/L{6\mu m}$ ratio.  
  Brusa et al. (2009b) reach similar conclusions by examining the 
   {\it Spitzer} colours of obscured AGN in the 1Ms CDFS observation.

\subsection{Optical Properties}

\begin{figure*}
  \resizebox{\hsize}{!}{\includegraphics{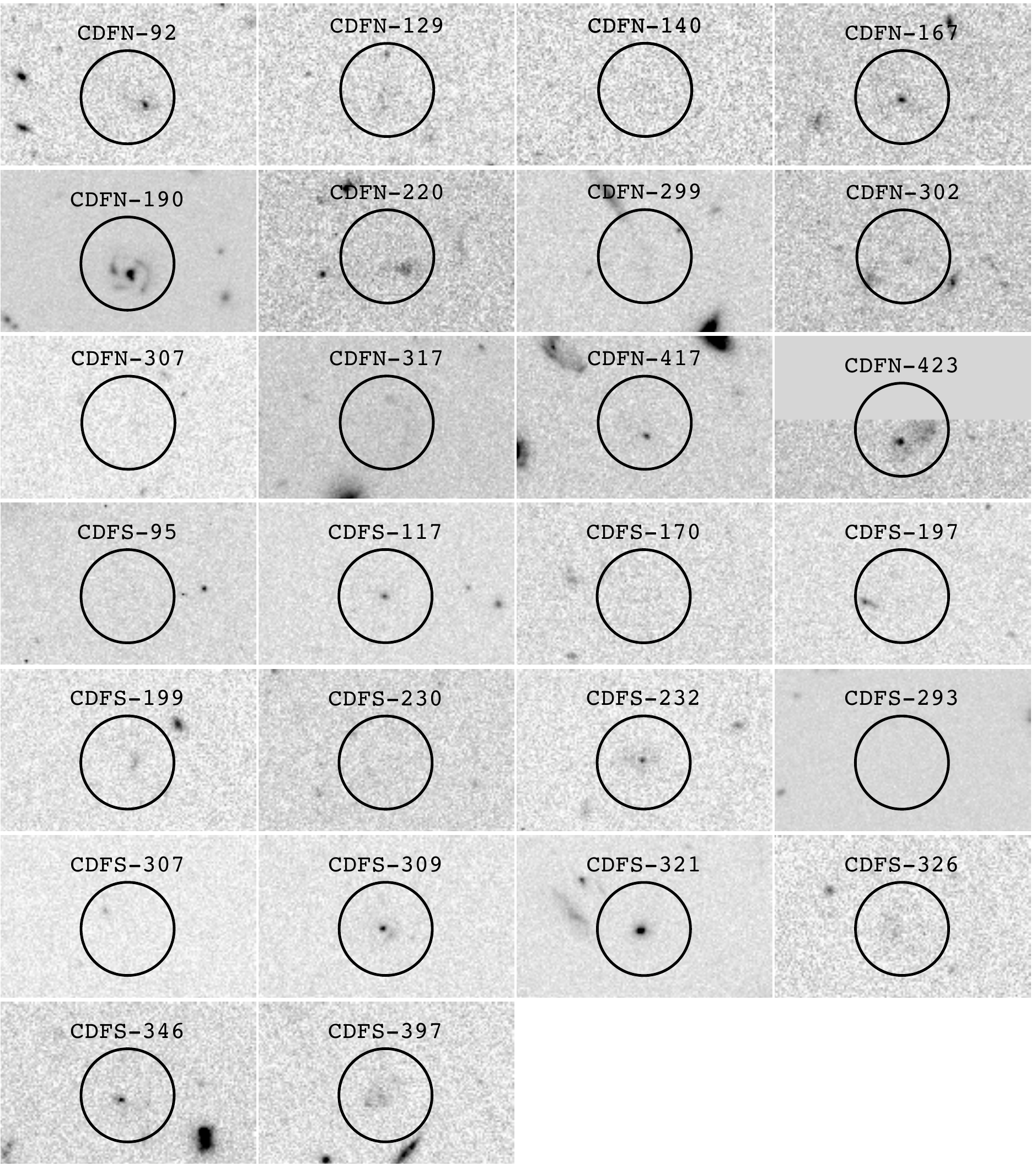}}
  \caption{Cutout images of the sources listed in Tab.\,\ref{sample}. In
           greyscale is shown the combined $B_{435}$-$V_{606}$-$I_{775}$-$z_{850}$
           image from the {\it HST} survey. Circles mark the positions of the
           MIPS 24$\mu m$ sources with 3\arcsec diameters}. Note: all {\it HST} counterparts 
            fall within the circle apart from the case of CDFS-95 which is the bright source westwards. 
  \label{cutouts}
\end{figure*}

In Fig.\,\ref{cutouts} we plot optical cutouts of the DOGs. The optical images
are the combined $B_{435}$-$V_{606}$-$I_{775}$-$z_{850}$ frames from the
{\it HST}-ACS survey in order to increase the signal-to-noise ratio. For CDF-S
sources there are {\it HST} detections for all the sources in Grazian et al. (2006).
In the north we are able to retrieve only 10/12 sources as there are no {\it HST} counterparts
for CDFN-140 and CDFN-307 in Giavalisco et al. (2004).  
In Table 6 we give the {\it HST} counterparts for the X-ray sources.  
For the detected sources we are able to 
inspect the morphology for many of them (typically when $z_{850}<26.5$). The
sources for which we are able to check the morphology are core-dominated but
there is a circumnuclear structure in a large fraction of them ($\sim 2/3$). 

The morphologies of the X-ray selected DOGs are qualitatively in line 
  with those of the overall DOG population. 
Melbourne et al. (2009) study with Keck adaptive optics the 
 near-IR images of 15 DOGs.  
 They find a correlation between the galaxy concentration
 and mid-IR luminosity in the sense that the most luminous 
 DOGs  exhibit higher concentration and smaller physical size.  
Bussmann et al. (2009a) present {\it HST} images of 31 DOGs.
 All but one of their DOGs (which follow AGN mid-IR SEDs) present spatially extended
 emission, but also 90\%  show significant unresolved components.

\begin{table*}
 \begin{center}
 \label{hstable}
 \caption{HST counterparts}
 \begin{tabular}{llllll}
 \hline 
  ID$^1$      &    HST$^2$  &  $\alpha$$^3$ & $\delta$$^3$   & z$^4$  & Morphology$^5$ \\
 \hline
CDFN-92 & 3844 & 189.04884 & 62.170823 & 25.42 & core+extended \\
CDFN-129 & 6339 & 189.09149 & 62.267672 & 27.84 &  \\
CDFN-140 & - & - & - & - &  \\
CDFN-167 & 8764 & 189.13037 & 62.166202 & 26.52 & core+extended \\
CDFN-190 & 10044 & 189.14828 & 62.240091 & 22.92 & spiral \\
CDFN-220 & 12070 & 189.17521 & 62.225481 & 25.97 & core+extended \\
CDFN-299 & 17100 & 189.24136 & 62.358056 & 27.18 &  \\
CDFN-302 & 17341 & 189.24465 & 62.249813 & 27.54 & core \\
CDFN-307 & - & - & - & - &  \\
CDFN-317 & 18291 & 189.25665 & 62.196300 & 27.28 &  \\
CDFN-417 & 25122 & 189.35646 & 62.285470 & 25.56 & core \\
CDFN-423 & 25391 & 189.36045 & 62.340789 & 24.61 & core+extended \\
CDFS-95 & 17814 & 53.034100 & -27.679634 & 25.25 & core \\
CDFS-117 & 11180 & 53.049049 & -27.774496 & 25.08 & core \\
CDFS-170 & 30122 & 53.071991 & -27.818924 & 26.27 &  \\
CDFS-197 & 30202 & 53.091526 & -27.853367 & 27.94 &  \\
CDFS-199 & 30084 & 53.092266 & -27.803137 & 26.21 & core+extended \\
CDFS-230 & 30199 & 53.105221 & -27.875071 & 27.14 & extended? \\
CDFS-232 & 15260 & 53.107006 & -27.718241 & 25.02 & core+extended \\
CDFS-293 & 30184 & 53.139332 & -27.874464 & 27.93 &  \\
CDFS-307 & 2818 & 53.146698 & -27.888338 & 26.06 & core \\
CDFS-309 & 7814 & 53.148827 & -27.821100 & 24.88 & core+extended \\
CDFS-321 & 4119 & 53.157345 & -27.870085 & 23.52 & core \\
CDFS-326 & 461 & 53.159641 & -27.931435 & 25.91 & extended? \\
CDFS-346 & 540 & 53.170151 & -27.929647 & 25.37 & core+extended \\
CDFS-397 & 1059 & 53.204872 & -27.917955 & 24.96 & core+extended \\
  \hline
 \end{tabular}
\begin{list}{}{}
\item The columns  are: (1) X-ray ID as in table 1; (2) HST ID from the catalogue 
 of Giavalisco et al. (2004) and Grazian et al. (2006) for the CDF-N and CDF-S respectively; 
 (3) {\it HST} coordinates (J2000); (4) z-band AB magnitude (F850LP)
 (5) comment on morphology.   
\end{list}
 \end{center}
 \end{table*}

\section{Discussion} 

\subsection{Overall Properties}
We investigate the properties of 26 X-ray selected DOGs in the 
 CDF-N and CDF-S.  Their  redshifts (at least for the 22  DOGs for which 
 there are available spectroscopic or photometric redshifts available)
 strongly cluster around  $z\sim 2.3$. 
Although the uncertainties are admittedly  high,
 as only four objects have a spectroscopic redshift available, 
 this finding comes in good agreement with the results of  
 Pope et al. (2008) and Dey et al. (2008).    
 Dey et al. (2008) measure spectroscopic redshifts for 
 86 DOGs in the Bootes field finding a redshift distribution
 centred at z=1.99. Pope et al. (2008) present mid-IR spectroscopy 
 for a sample of 12 (out of 79)  Spitzer-selected DOGs in GOODS-N, finding 
 that the spectra lie in a tight distribution around $z\sim 2 $. 
   We find that all our sources have $L_{IR} > 10^{12} L_\odot$ 
 and thus are classified as ULIRGs. The uncertainties in the estimations 
 of the total IR luminosities, based on only IRAC and MIPS data,
  are expected to be within a factor of 3
 (see Dey et al. 2008, Bussmann et al. 2009b). This has to do with the 
 fact that there are no far-IR or sub-mm measurements available
 (apart from one source) and thus the amount of cold 
 dust available in these systems is uncertain. 
 Bussman et al. (2009b) using SHARC-II 350 $\mu m$ imaging 
 of 12 DOGs in the Bootes field, estimate 
 IR luminosities $2\times 10^{13} L_\odot$, higher than 
 other classes of dusty $z\sim2$ galaxy populations such 
 as sub-mm galaxies. 

  The vast majority of our sources are most probably hosting an AGN 
 owing to the large X-ray luminosities ($\rm L_{2-10}>3\times 10^{42}$ \lunits)
  and high $L_X/L_{6\mu m}$ luminosity ratios (with the possible exception of 
   the sources CDFN-317 and CDFS-309 see the discussion in section 6.2). 
However,  the mid-IR SEDs fail to reveal the presence of an 
 AGN in many cases, suggesting that the AGN does not dominate 
  the IR bolometric output.  
 Only seven sources follow a power-law flux 
 distribution suggestive of a broad range of warm dust temperatures
 similar to that observed in AGN dominated sources.
 The Mkn231 ULIRG template provides a good representation of these SEDs.
 Bussmann et al. (2009b) also find the same for their 
 DOGs with sub-mm observations.   
 The majority of our sources show a bump in their flux distribution which 
 is suggestive of a stellar continuum peaking at rest-frame 
 wavelength of 1.6$\mu $. 
 These SEDs are well represented by the Arp220 ULIRG template.
 The mixture of star-forming and AGN SEDs we observe  
 in our X-ray selected DOGs is quite similar to that observed 
 by Dey et al. (2008) in their {Spitzer}-selected sample. 
 We note that the mean luminosity of the sample of Dey et al. (2008) 
  is  $\log [\nu L_\nu (8\mu m)]= 12 $  
   ($\sigma=0.64$) while in our sample the luminosities are   somewhat lower
    with  $\log [\nu L_\nu (6\mu m)]= 12 $ ($\sigma=1$). 

 \subsection{Comparison with other X-ray observations of DOGs}
Lanzuisi et al. (2009) have recently performed an X-ray
study of  luminous  DOGs ($F_{24 \mu m}/F_R>2000$ 
 and $F_{24 \mu m} >1.3$ mJy ) in the Spitzer Wide-Area  
 Infrared Extragalactic (SWIRE) survey. 
 An area of 6$\rm deg^2$ out of the total 50$\rm deg^2$  
  of the SWIRE survey has been surveyed by {\it XMM-Newton} 
 or {\it Chandra}. Their sample includes 44 DOGs 
 in the redshift range $0.7<z<2.5$ of which 
 23 are detected in X-rays. These DOGs appear to be highly 
 absorbed with about 50\% having a column density 
 $\rm N_H>10^{23}$ \cunits while one source is associated 
 with a transmission dominated Compton-thick QSO.  
 In comparison, in our sample out of the 18 objects with 
 available redshifts and in which we were able to perform spectral fits, 
 11 have $N_H> 10^{23}$ \cunits. 
  It appears that despite the fact that the SWIRE sample 
   is much brighter (its median flux is $\sim1\times10^{-13}$ \funits),
   the fraction of highly absorbed sources ($>10^{23}$ \cunits) is not very different 
    compared to our sample.
    However, the fraction of candidate Compton-thick sources is 
     substantially higher in our fainter sample. 
     
     Fiore et al. (2009) investigate the X-ray properties of bright 24$\mu m$ sources 
      in the COSMOS survey (Elvis et al. 2009). 
     They select 73 DOGs of which 31 are detected in X-rays having a mean
      luminosity of $\rm log[L_(2-10 keV)] \approx 43.5$ at a mean    
      redshift z=1.55. The hardness ratio of these sources is  $0.50\pm 0.34$ 
       in the 0.3-1.5 vs. 1.5-6 keV bands corresponding to 
        a photon index of $\Gamma\sim 0.5$. This hardness ratio is 
         identical to that of the remaining 42 non-detected DOGs (0.53$\pm0.14$).
          The   spectrum of the COSMOS DOGs is harder but consistent within the errors 
           with the derived average spectrum of the DOGs in or sample.     
     
     Georgakakis et al. (2010) follow a different approach for identifying DOGs. 
      They identify lower (spectroscopic) redshift $z\approx 1 $ sources 
       in the AEGIS and CDFN surveys  which are analogues to the distant  DOG population
        at $z \sim2$. That is  their SED are similar to these of DOGs so
 if placed at redshift of z=2 they would classify as DOGs.  
        They find nine such sources with X-ray counterparts. Their X-ray spectra are consistent with 
          Compton-thin obscuration with only three sources presenting tentative
           evidence for Compton-thick obscuration.
            Ten more Infrared-excess sources have no X-ray counterparts. 
            Their SEDs are consistent with starburst activity showing 
            no evidence for a hot dust component.    
    Georgakakis et al. (2010) conclude that there is little evidence for 
      the presence of a large fraction of luminous Compton-thick sources 
       in either the X-ray detected or undetected  population of DOG analogues. 
       As far as the X-ray detections are concerned, our findings  
        are not that discrepant quantitatively,  
         given that our fraction of Compton-thick sources is  
        4/12 among the brightest  sources.  
       
\subsection{Comparison with X-ray undetected DOGs}
 
 Fiore et al. (2008) examine the properties of DOGs in the CDF-S area
  which are undetected in the X-ray image. 
  They find that the SEDs of most of these DOGs are dominated by AGN 
   emission. The stacking analysis yields a hardness ratio 
    corresponding to a flat spectral index of $\Gamma\sim1$.
     They conclude that $80\pm15$ \% of these are likely 
      to be highly obscured, Compton-thick AGN.  
     Georgantopoulos et al. (2008) have performed the same 
      exercise in the Chandra Deep Field North finding 
       a similar hardness ratio corresponding to a 
        spectral index for the undetected DOG ($\Gamma \approx0.8$).
  The average flux of the undetected DOGs in the CDF-N 
 is $\rm f_{2-10 keV}\sim 6\times10^{-17}$ \funits (as compared to 
  $\rm f_{2-10 keV}\sim 9\times 10^{-16}$ \funits for the detected ones in this work).
        Fiore et al. (2009) look for X-ray undetected DOGs in the 
         Chandra-COSMOS survey (Elvis et al. 2009).
          They claim that the number density of Compton-thick QSOs 
           associated with DOGs is about half of all X-ray selected QSOs.    
  Finally, Treister et al. (2009b) examine the properties of 
 211 heavily obscured AGN candidates in the Extended CDF-S selecting 
 objects with $f_{24\mu m}/f_R > 1000$ and $R-K >4.5$. 
 18 sources are detected in X-rays having moderate column densities $\sim 10^{22-23}$ \cunits
 while two of them appear to be Compton-thick. On the other hand 
 the undetected sources show a hard average spectrum which could
 be interpreted as a mixture of 90\% Compton-thick objects and 
 10\% star-forming galaxies.      
         It should be kept in mind though that 
  as only an average hardness ratio can be derived in the stacking analysis 
    one cannot definitively 
   rule out the possibility that some of the sources are moderately 
    absorbed ($\rm\sim 10^{23}$ \cunits at z=2) instead of intrinsically flat. 
   Our work here, which examines the individual X-ray spectra of 
   the (albeit brighter) X-ray detected DOGs can provide an answer to this question.  
  We find a mixed bag of objects. 
 At least  four out of the 22 sources with derived spectra 
  are good candidates for being reflection dominated Compton-thick sources.
   At most 12 sources could be intrinsically flat and thus 
    Compton-thick candidates but unfortunately the 
     limited photon statistics do not allow us to 
      discern  between  an intrinsically flat spectrum 
       or a moderately absorbed one.
  At the same time a large number of DOGs 
   present relatively steep spectra ($\Gamma > 1.4$) and thus
    are unambiguously associated with unabsorbed  sources.    
    One could argue that the X-ray detected DOGs 
     are expected to be less absorbed compared to the undetected ones.  
 This is simply because larger column densities diminish 
  the X-ray flux making the sources fainter than the CDF X-ray flux limit. 
  Yet it is interesting that the detected DOGs have an 
   X-ray spectrum very similar to the stacked spectrum of undetected DOGs 
    ($\Gamma \sim 1$) or flatter. Since a significant  number of
     the X-ray detected DOGs in the CDFs are showing moderate absorption, 
      it is not unlikely that undetected DOGs follow a similar absorption pattern.    
 Nevertheless there is a significant difference between the X-ray detected 
 and the undetected DOGs. The former have an X-ray to mid-IR luminosity ratio 
 of  $\rm L_{2-10 keV}/L_{6\mu m}\approx -1.3 $ (median) while the latter 
 have   $\rm L_{2-10 keV}/L_{6\mu m}\approx -3 $ (see Fig.2 of Georgakakis et al. 2010).
 This could imply that the samples of undetected DOGs indeed contain a large 
 fraction of Compton-thick sources. Then the fact that the 
 average spectrum of the X-ray detected DOGs is comparable to that 
 of undetected DOGs is a mere coincidence.   
Alternatively, there may be a large contamination of the X-ray undetected DOG 
 samples from normal galaxies. Galaxies in the local Uninerse  are known to 
have low X-ray to mid-IR luminosity ratios (Ranalli et al. 2003).   

\section{Summary}
 The mid-IR excess techniques 
 have attracted much attention for claiming to be able to select Compton-thick AGN
  in a very efficient way. 
  Here, we  present a sample of 26 X-ray detected infrared-excess sources
   (DOGs) in the Chandra Deep Fields having $f_{24\mu m}/f_R>1000$. We 
 examine their properties attempting to address the Compton-thick content 
 of this sample mainly using X-ray spectroscopy.
    Our conclusions can be summarised as follows. 
   \begin{itemize}  
    \item{We find that an appreciable  fraction 
       of the DOGs (7 out of 22)  are associated with relatively unobscured sources,  
   having $\Gamma > 1.4$. These are unobscured AGN rather than normal galaxies 
    as confirmed by their high X-ray luminosities.}
    \item{ We identify a number of intrinsically flat $\Gamma <1 $ sources:
     4 out of the 12 with good photon statistics.  
    These could be associated with Compton-thick AGN. If we include 
     the sources with poorer photon statistics, 
     the number of flat sources could be as many as 12 out of 26.
       In this respect we find that the infrared-excess techniques can indeed 
 unveil a significant number of heavily obscured AGN at high redshifts.} 
 \item{The fraction of Compton-thick sources could then 
 vary roughly from  30\% to 50\%. This is clearly lower than  
 reported in previous studies of stacking analysis of X-ray 
 undetected DOG samples.}
 \item{The average spectrum of the X-ray detected DOGs is very similar   
  if not  harder than that of undetected DOGs in the Chandra Deep Fields.} 
 \end{itemize}

\begin{acknowledgements}  
We would like to thank the referee Fabrizio Fiore for many useful comments 
 and suggestions. 
IG and AC acknowledge the Marie Curie fellowship
 FP7-PEOPLE-IEF-2008 Prop. 235285.
 AC acknowledges receipt of ASI grants I/023/05/00 and 1/88/06. 
We acknowledge the use of {\it Spitzer} data provided by the 
 {\it Spitzer} Science Center. 
 The Chandra data were taken from the Chandra Data Archive 
 at the Chandra X-ray Center.   
\end{acknowledgements}

\end{document}